\documentclass{mcmthesis}
\mcmsetup{CTeX = false,   
        tcn = \red{2226491}, problem = \red{C},
        sheet = true, titleinsheet = true, keywordsinsheet = true,
        titlepage = false,abstract =true}
\usepackage{newtxtext}
\usepackage{lipsum}
\usepackage{amsmath}
\usepackage{amssymb}
\usepackage{graphicx}
\usepackage{float}
\usepackage{tabu}
\usepackage{color}

\newcommand{\red}[1]{\textcolor[rgb]{1,0,0}{#1}}

\newcommand{\tabicell}[2]{\begin{tabular}{@{}#1@{}}#2\end{tabular}}
\title{Trading Strategies: Earning More in Investment}
\begin{document}
\begin{abstract}

Gold and bitcoin are not new to us, but with limited cash and time, given only the past stream of daily price of the gold and bitcoin, it is a kind of new problem for us to develop a certain model and determine a best strategy to get the most return. 

Here, we team members analyzed the data provided and finally made a unified system of models to predict the price and evaluate the risk and return in our act of investment, and we name this series of model and measurements as $'{CTP} {Model}'$. This is a model which can determine and describe what transaction should the trader make on each day and what is the certain maximum return he will get under different risk levels.

Firstly, considering the relatively high price for the gold and bitcoin, which may easily diverge in the process of calculation, we first normalize the everyday investment status to form dynamic programming model. By assuming the proportion of funds applied on the investment of certain item in the total value, we symbolize different investment strategies in a mathematical formula and its restrictions. Secondly, we come to the process of prediction model, where we first did the preprocessing step on the data by the method of piecewise Hermite Interpolation, for bitcoin can be traded every day while gold can only be traded on weekdays. In the next step, we applied one of the time series model--$\mathbf{ARIMA}$ model, with which we can make predictions on the day of investment. From the analysis of ACF and PACF, we can determine the final model to be $\mathbf{ARIMA}(1,1,1)$. Due to the large fluctuations of prices of gold and bitcoin in the real stock market, we further optimize the $\mathbf{ARIMA}$ model by selecting and controlling the amount of past data used. Continuously adjusting the time length $T$ to find the minimum value of the coefficient of determination $R^2_{min}$, finally we got the optimal time length 60. Thirdly, we come to the Investment Risk and Return Model with the construction of Markowitz Mean-Variance Model and Sharpe Ratio Investment Model Based on Particle Swarm Optimization. For the former one, we use it to find the ratio of optimal asset allocation. But since it is basically based on the assumption that gold and bitcoin are relatively independent, which is hard to achieve in the real world, we turn to the Sharpe ratio model which better suits the real investment world. When the target function of Sharpe ratio model reaches its largest number, that is the balance state of risk and return. To help better speed up the convergence, we applied the Particle Swarm Optimization algorithm(PSO) and eventually get the three types of maximum return with different target function, that is to say, risk levels.

Considering adding a certain disturbance to the results of our investment strategies, it can be seen from the results that the final result after perturbation is not as good as the best result taken by our model, so the solution obtained by our model is the optimal solution.

In our testing of sensitivity, the model has corresponding changes to the disturbance of the original total assets and transaction costs and then embodies more traits in inclination in investment of bitcoin and gold in the model with the disturbance.

\end{abstract}
\maketitle

\tableofcontents
\newpage
\section{Introduction}
    In order to better present the problem and our solutions as well as the process of modeling, the following background and information is worth mentioning.
\subsection{Literature Review}
    As is known to all, market traders will buy and sell volatile assets frequently to maximize their total return. The two main assets they process with are gold and bitcoin and within their each purchase and sell, there will always be a commission. Given past stream of daily prices of gold and bitcoin, we are asked to develop a model to determine each day if the trader should buy, hold, or sell their assets in their portfolio.
    
    On each trading day, the trader will have a portfolio consisting of cash, gold and bitcoin and during each day transaction, there will also be commission which cost differently for gold and bitcoin. Starting with \$1000 and given 5-year period of trading time, the trader can trade bitcoin everyday but only weekdays for gold. We are asked to build a model to forecast and arrange what the trader should buy, hold, or sell his assets in portfolio to get a maximum return.
    
\subsection{Restatement of the Tasks}
    In order to clarify our tasks, we list our tasks below:
    
    Task 1: Based on the data of price on exactly that day, the model we develop should offer the best daily trading strategy, and come up with the final result that how much is the initial \$1000 worth at the end of the 5 years period of time with our model and strategy.
    
    Task 2: Include the evidence to prove that the model we develop and apply provides the best strategy of processing assets every day.
    
    Task 3: Taking the cost of transaction into account, we are obliged to work our the sensitivity of the strategy to the transaction costs, including to what extent will the transaction costs affect the strategy and the results.
    
    Task 4: Write down all the strategy, model and results in a memorandum of at most two pages to make the trader understand and know how to act with the model.
    
\subsection{Interpretation of the Phrases}
Time Series Model: In industry and scientific research, the observation and measurement of a certain or a group of variables $x(t)$ will be arranged in time order at a series of moments $t_1$, $t_2$, …, $t_n$ and used for explanatory variables and mathematical expressions of interrelationships.

Asset allocation: The main problem to be settled in the asset allocation is that how to diversify investment to maximize returns while minimizing risk. The main goal of asset allocation is to achieve a certain expected return target at a certain point in the future, and to control the fluctuation of assets within a certain range that an individual can accept.

Sharpe ratio: Sharpe ratio not only pays attention to the return of the asset, but also pays attention to the risk of the asset. It measures the return of the asset after adjusting the risk, and display the price of the unit risk. Since the Sharpe ratio comprehensively reflects the risk-return characteristics of the capital market, it has been widely used in evaluating the performance of asset portfolios, judging the operating efficiency of the capital market, constructing effective asset portfolios, and guiding investment decisions.

Particle Swarm Optimization algorithm: It is a swarm intelligence algorithm that simulates the mutual cooperation mechanism of the foraging behavior of biological groups in nature to find the optimal solution to the problem. The algorithm has the advantages of simple principle, easy implementation and less control parameters. The basic idea is to randomly select a group of particles in the solution space and randomly distribute them to the solution space. The speed and direction of each particle determine the next position of the particle. The historical optimal solution found by the particle itself and the historical optimal solution found by the entire group affects the movement speed and direction of each particle in the next turn. Each particle is regarded as a feasible solution of the objective function, and the position value of the particle is brought into the fitness function to calculate and evaluate the quality of the solution and finally get the global optimal solution. 

\section{Symbol Description}

    \begin{table}[H]
        \centering
        \begin{tabular}{cl}
            \\[-2mm]
            \hline
            \hline\\[-2mm]
            {\bf Symbol} & \qquad {\bf Meaning}\\
            \hline
            \vspace{1mm} \\ [-3mm]
            $S_i = [C_i,G_i,B_i]$ & \tabicell{l}{The investment status before the transaction on day $i$, which is the \\ amount of cash $(C)$, gold $(G)$ and Bitcoin $(B)$ owned by the trader.} \\
            \vspace{2mm}
            $P^g_i$ & The price of gold on day $i$. \\
            \vspace{2mm}
            $P^b_i$ & The price of the Bitcoin on day $i$. \\
            \vspace{2mm}
            $s_i = [c_i,g_i,b_i]$ & The investment status before the transaction on day $i$ after normalization. \\
            $T$ & \tabicell{l}{The amount of data used by the prediction model $\mathbf{ARIMA}$. For example, \\on the $13^{th}$ of September 2017, if we use $T=60$ to forecast, the data \\ volumn is the data of 60 days from July $15^{th}$, 2017 to September $13^{th}$, \\ 2017.}\\
            $SR_L$ & The sharpe ratio. \\
            $\mu(i)$ & \tabicell{l}{The target function of the dynamic programming model.}\\
            \hline
            \hline
        \end{tabular}
    \end{table}
    
\section{The Setup of the Model} \label{The Setup of the model}
    In order to acquire the best investment strategy to maximize the profits of investment, we should make a refreshment of the investment state $S =[C,G,B]$, which is dynamic programming every day. We can have different ways to invest for each day, but by analyzing the gold and bitcoin data before the day to predict the price changes in the next few days, we can eventually come up with the best way to invest.
    \begin{figure}[H]
        \small
        \centering
        \includegraphics[width=\textwidth]{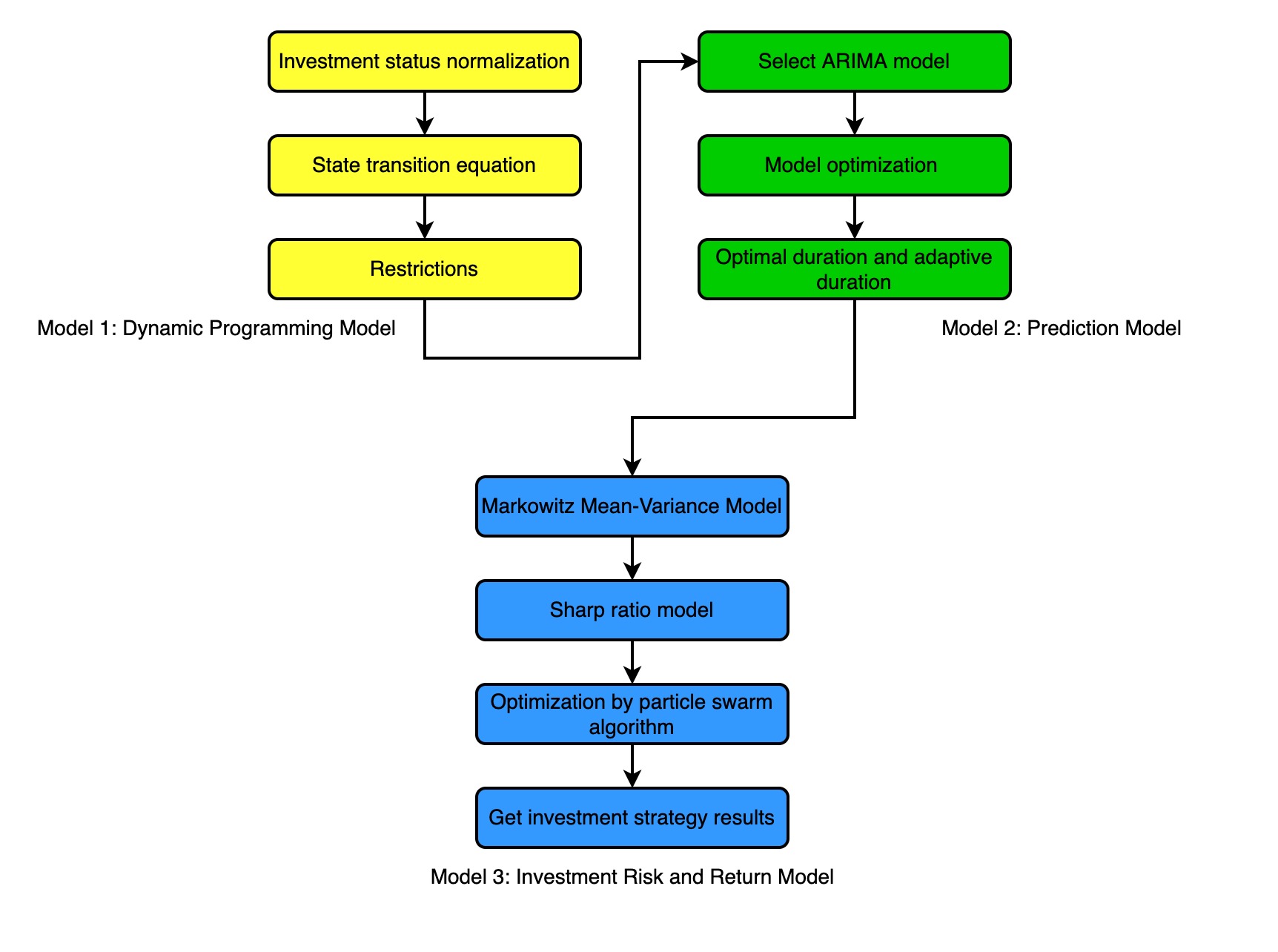}
        \caption{Model Overview} \label{fig:ALl}
    \end{figure}
\subsection{Dynamic Programming Model}
    Because of the relatively high prices of gold and bitcoin, it is quite easy to meet divergence in the process of calculation. So we choose to normalize the everyday investment status.
    
    Supposed that the overall value on the $i^{th}$ day is $V_i$, which is:
    \begin{align}
        V_i = C_i + G_i \times P^g_i + B_i \times P^b_i
    \end{align}
    define the normalized investment status $s_i = [c_i,g_i,b_i]$ to be:
    \begin{align} 
        & c_i = C_i /  V_i  \\
        & g_i = G_i  P^g_i / V_i  \notag\\
        & b_i = B_i  P^b_i / V_i \notag
    \end{align}
    After such operation, it can be guaranteed that
    \begin{align}
        c_i + g_i + b_i = 1
    \end{align}
    
    For example, the normalized investment status before the $i^{th}$ day is $s_i = [c_i,g_i,b_i]$.Trade gold $\Delta G_i$ on that day, buy bitcoin $\Delta B_i$, considering that transactions incur commission costs($\alpha = 1\%$ and $\beta = 2\%$ respectively), the amount of change in cash is
    \begin{align}
        \Delta C_i = - P^g_i \Delta G_i(1 + \alpha) - P^b_i\Delta B_i(1 + \beta) \label{Delta_G}
    \end{align}
    Then before the $i+1^{th}$ day's transactions, due to the change in price of gold and bitcoin, the change in the normalized investment status is
    \begin{align}
       & c_{i+1} = (C_i + \Delta C_i ) / V_{i+1} \label{Delta_s}\\
       & g_{i+1} = (G_i + \Delta G_i) P^g_{i+1} / V_{i+1} \notag \\
       & b_{i+1} = (B_i + \Delta B_i) P^b_{i+1} / V_{i+1} \notag
    \end{align}
    Let the proportion of funds used to invest in a certain item in the total value in one day be
    \begin{align}
        & x =  P^g_i \Delta G_i / V_i \\
        & y =  P^b_i \Delta B_i / V_i \notag
    \end{align}
    From \ref{Delta_G} and \ref{Delta_s}, we can use $x,y$ to indicate the change in total funding
    \begin{align}
        V_{i+1} = V_i [1 + (\frac{P_{i+1}^g}{P_i^g} - 1 - \alpha)x + (\frac{P_{i+1}^b}{P_i^b} - 1 - \beta)y + (\frac{P_{i+1}^g}{P_i^g} - 1)g_i + (\frac{P_{i+1}^b}{P_i^b} - 1 )b_i]
    \end{align}
    In the same way, we can also use $x$ and $y$ to indicate $s_{i+1}$
    \begin{align}
        & c_{i+1} = \frac{V_{i}}{V_{i+1}}[c_i - x(1 + \alpha) - y(1 + \beta)] \\
        & g_{i+1} = \frac{V_{i}}{V_{i+1}} \frac{P_{i+1}^g}{P_i^g} (x + g_i) \notag \\
        & b_{i+1} = \frac{V_{i}}{V_{i+1}} \frac{P_{i+1}^b}{P_i^b} (y + b_i) \notag
    \end{align}
    That means we can choose different values for $x$ and $y$ to symbolize different investment strategy. Since none of the three terms in the investment state for each day can be less than zero,  the natural constraints of the model are
    \begin{align}
        & c_i - x(1 + \alpha) - y(1 + \beta) \geq 0 \\
        & x + g_i \geq 0 \notag \\
        & y + b_i \geq 0 \notag \\
        & x = 0 \ \mathrm{For \ weekend} \notag
    \end{align}
    
    In order to acquire a better investment strategy, that is, to choose the optimal value for $x$ and $y$ to achieve the desired return every day, we need to predict the price changes after the day of the investment, and also optimize the dynamic programming model according to the relationship between investment risk and return.

\subsection{Prediction Model}
\subsubsection{Preprocessing of Data}
    To make our prediction model more accurate, we find the price change of gold from 1968 to 2016 and the price change of bitcoin from 2010 to 2016 respectively. Since gold is only traded on the opening day, we only know the price data of $5$ days from Monday to Friday during the week, which is very inconvenient for the forecasting process. So in this essay, we apply the method of piecewise Hermite interpolation to approximate the fit of gold price data on Saturday and Sunday from 1968 to 2016. For the Saturday and Sunday after September $11^{th}$, 2016, the gold price is estimated by the method of forecasting.
\subsubsection{Time Series Model:\textbf{ARIMA} model}

    By the means of predicting future data, it can help us better choose the optimal investment strategy. The price of gold and bitcoin always exists certain trend which can generally be stabilized after gradual differentiation. Then we can apply the $\mathbf{ARIMA}(q,p,d)$ model. So we can use the $\mathbf{ARIMA}$ model to make predictions on the day of investment.
    
\begin{figure}[H]
        \small
        \centering
        \includegraphics[width=0.9\textwidth]{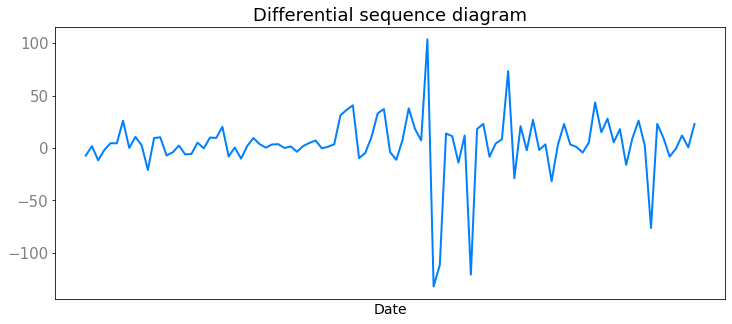}
        \caption{The first-order difference diagram} \label{fig:dsd}
    \end{figure}

\begin{figure}[H]
        \small
        \centering
        \includegraphics[width=0.9\textwidth]{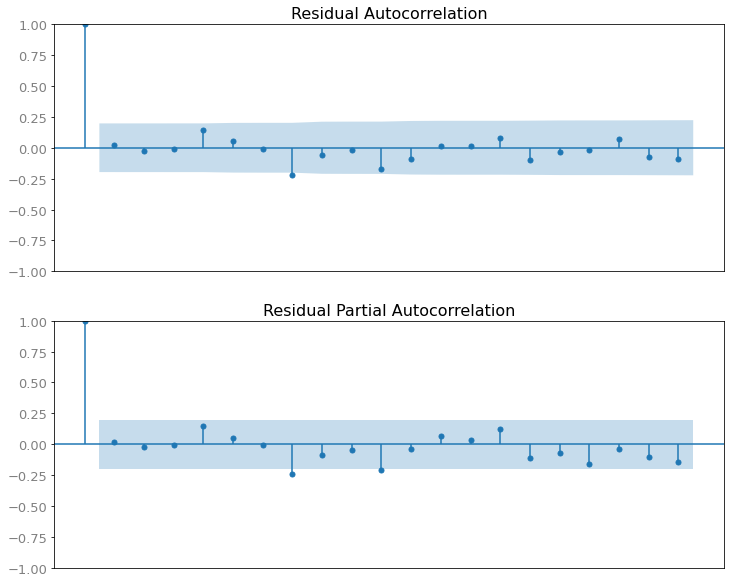}
        \caption{RACF and RPACF} \label{fig:RACF_PRACF}
    \end{figure}

    Let $Z_t$ be the stationary sequence of $Y_t$ after the $d$-order difference operation, that is:
    \begin{align}
        Z_t=\nabla^dY_t\left(t>d\right)
     \end{align}
    $Z_t$ is the form of $\mathbf{ARMA}(p,q)$, and $Y_t$ is called $d$-order summation sequence of $\mathbf{ARMA}$. The general form of the model $\mathbf{ARMA}\left(p,d,q\right)$ is:
    \begin{align}
        \phi\left(B\right)\left(1-B\right)^dY_t=\theta\left(B\right)e_t
    \end{align}
    The establishment process is shown in Figure \ref{fig:ARIMA_process}:
    \begin{figure}[H]
        \small
        \centering
        \includegraphics[width=\textwidth]{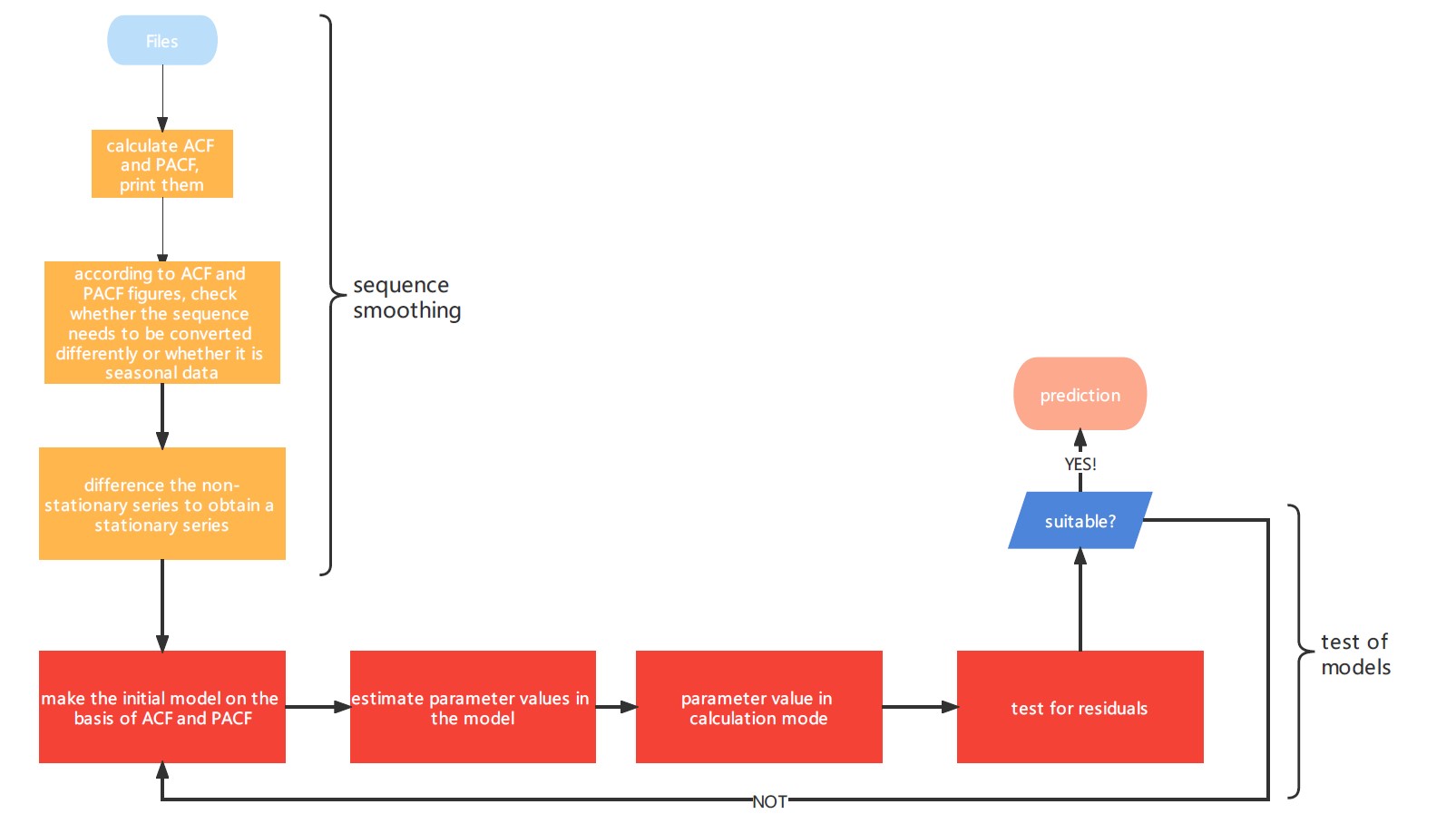}
        \caption{The process of ARIMA} \label{fig:ARIMA_process}
    \end{figure}
    For example, on February $17{th}$, 2017, the bitcoin price fluctuation data of 100 days before that day was used to predict the bitcoin price change for the next three days. After performing the first-order difference on the data as shown in Figure \ref{fig:dsd}, we can find that the data has become stable after the first-order difference, so we can make the parameter $d = 1$ in the model.
    After the first-order difference, the data has been in the state of stabilization, so we can make the parameter $d = 1$ in the model. After that, we perform autocorrelation analysis on the data after the first-order difference, and draw its autocorrelation and partial autocorrelation. The correlation diagram is shown in the figure \ref{fig:ACF_PACF}:
    \begin{figure}[H]
        \small
        \centering
        \includegraphics[width=0.9\textwidth]{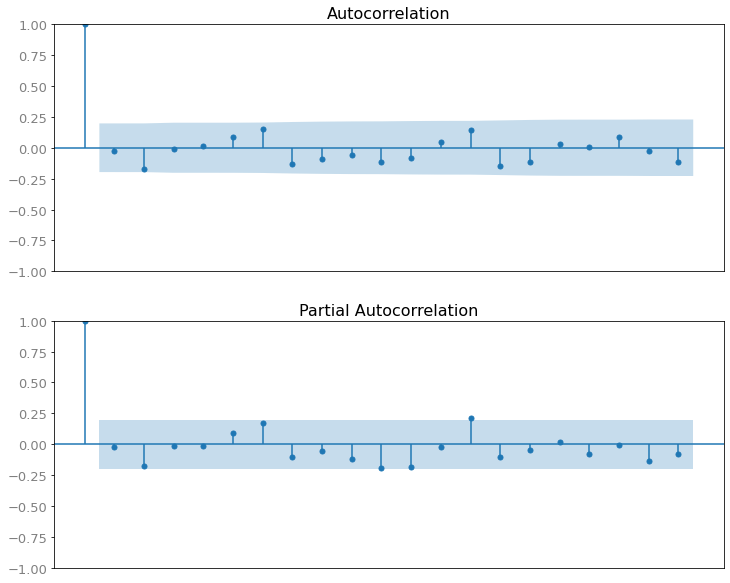}
        \caption{ACF and PACF} \label{fig:ACF_PACF}
    \end{figure}
    From the analysis of ACF and PACF, it can be known to all that the parameters in the model can be taken as $p = 1$ and $q = 1$, and then we can get the final model $\mathbf{ARIMA}(1,1,1)$. Fitting with this model can be obtained as shown in Figure \ref{fig:od_fr}:
    \begin{figure}[H]
        \small
        \centering
        \includegraphics[width=0.9\textwidth]{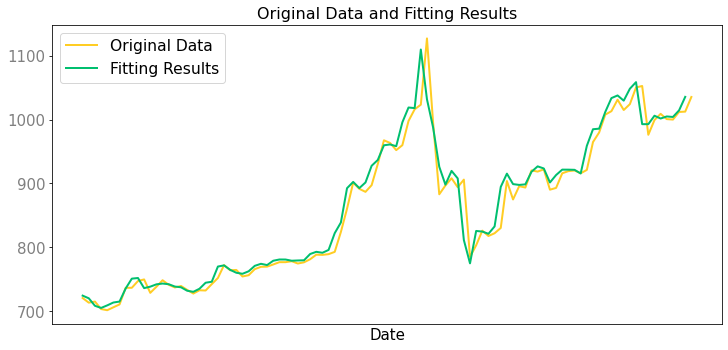}
        \caption{Original data and its fitting results by the model} \label{fig:od_fr}
    \end{figure}
    At last, in order to prove that the residuals in this model do not have autocorrelation, that is, the model residuals are white noise, the autocorrelation analysis of the residuals is shown in Figure \ref{fig:RACF_PRACF}. It can be seen that the residual of the prediction result of the model is a white noise sequence, which meets the requirements of the $\mathbf{ARIMA}$ model.
\subsubsection{Model optimization}
    From the above process of data, we can find that although the prediction results of the prediction model are in good agreement with the actual results, large errors may still occur. This is because in our real life, the prices of gold, bitcoin and other stock markets are often subject to human factors (such as fluctuations due to policies and news) and other factors, resulting in unstable terms which will make the prediction effect of $\mathbf{ARIMA}$ model become weak. In order to solve this problem, we choose to control the amount of past data used for the $\mathbf{ARIMA}$ model.

    We select a period of time $T$ which is suitable enough to predict the price changes in the next three days. This sufficient period of time needs to satisfy two conditions:
    \begin{enumerate}
        \item The time length $T$ contains enough price data for the model to fit well.
        \item The time length $T$ should not be particularly long. In the data within this time length, we need to ensure that the influence of unstable items caused by various factors is small enough.
    \end{enumerate}
    So here we use the coefficient of determination:
    \begin{align}
        R^2 = \frac{\mathbf{SSR}}{\mathbf{SST}} = \frac{\sum_i^n (\hat{y}_i - \bar{y})}{\sum_i^n (y_i - \bar{y})}
    \end{align}
    to characterize the fitting effect of the model in a certain time length $T$.
    
    For this situation, we use a combination of optimal time length and adaptive time length. Here, for the accuracy of the model, we find the price change of gold from 1968 to 2016 and the price change of Bitcoin from 2010 to 2016.
    
\begin{figure}[H]
        \small
        \centering
        \includegraphics[width=0.9\textwidth]{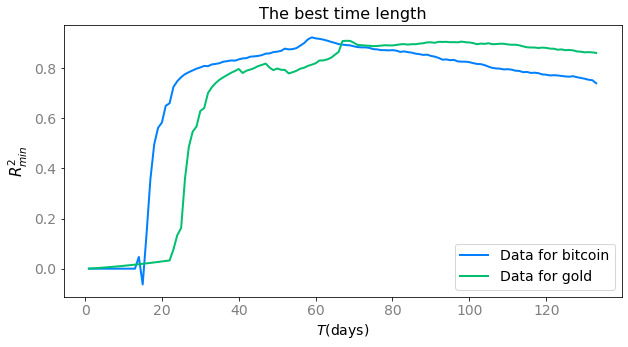}
        \caption{The best time length of $\mathbf{ARIMA}$ model} \label{fig:Best_T}
    \end{figure}

    The process of finding the best time length is to use the same time length to fit the data for the every last day, and calculate the smallest coefficient of determination $R^2_{min}$ in all fitting results, continuously adjust the time length $T$ to find the minimum value of $R^2_{min}$, and finally get the best time length.The result is shown in the figure \ref{fig:Best_T}. It can be observed that when $T = 60$ is selected, the forecast fitting effect of the $\mathbf{ARIMA}$ model on gold and bitcoin prices is very good, so the optimal time length is $60$.
    
\begin{figure}[H]
        \small
        \centering
        \includegraphics[width=\textwidth]{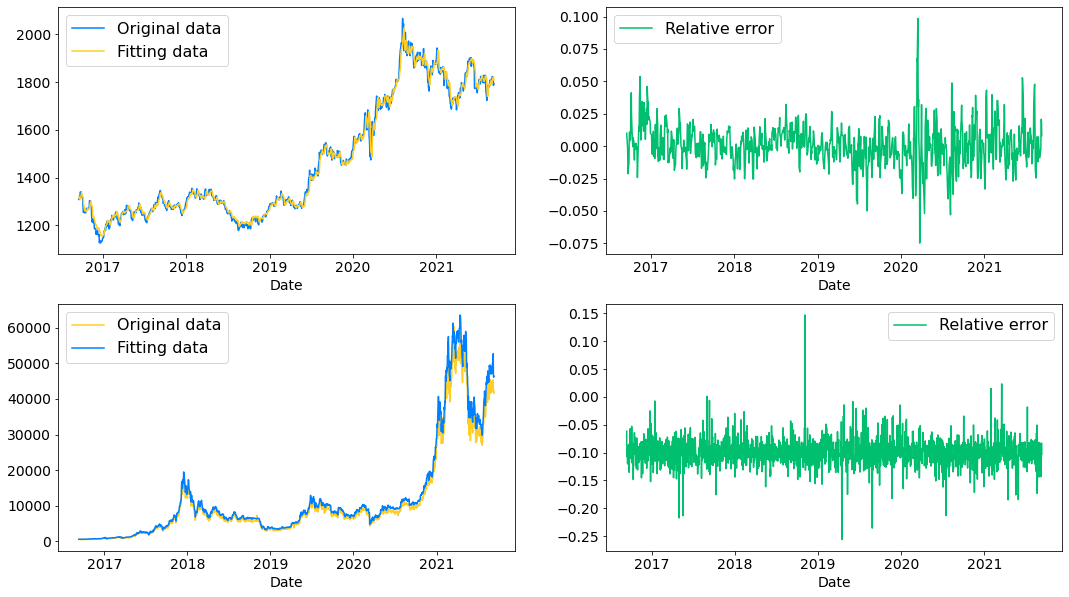}
        \caption{Prediction data result and relative error} \label{fig:pred}
    \end{figure}

    Since the optimal time lenth mentioned above is the measurement of the overall data, there may be a more suitable time length for some specific data, so it needs to be self-adapted according to different data. The adaptive scheme we choose starts from the optimal time length $T=60$, and uses $T-1$ and $T+1$ for prediction in the predicting process, judges the coefficient of determination of the three, and then selects the largest time length corresponding to the coefficient as the new $T$. The above steps are continued and repeated until the maximum corresponding determination coefficient $R^2$ is found, that is, the point with the best fitting effect.The results of prediction int this method are showen in figure \ref{fig:pred}.
\subsection{Investment Risk and Return Model}
\subsubsection{Markowitz Mean-Variance Model}

    The Markowitz mean-variance model is used to find the ratio of optimal asset allocation. Assuming that there are $n$ kinds of risky assets in the market and the returns of the assets are $y_1, y_2, \dots, y_n$ respectively. The allocation ratio of investors in each risky asset is $e_i$ for each, so the return rate of the portfolio is is $y_p = \sum_i^n e_i y_i$, in which:
    \begin{align}
        \sum_i^n e_i = 1
    \end{align}
    So the expected return rate of the portfolio is: 
    \begin{align}
        E(y_p) = \sum_i^n e_i E(y_i)
    \end{align}
    The variance is:
    \begin{align}
        D(y_p) = \sum_{i=1}^n e_i^2 D(y_i) + \sum_{i\neq} e_i e_j \mathbf{Cov}(y_i,y_j)
    \end{align}
    The precondition of investors' asset allocation is to form expectations for the future, that is, the probability distribution of $y_i$. then they can set their own expected return targets, and finally determine the investment ratio $\omega_i$ in each risky asset to achieve investment goals.
    Supposing that the investor's initial asset is $V_0$, so the future asset is $V_0(1 + y_p)$. According to our description in the subsection \ref{The Setup of the model}, we have:
    \begin{align}
        1 + y_p = \prod_{i}^k [1 + (\frac{P_{i+1}^g}{P_i^g} - 1 - \alpha)x + (\frac{P_{i+1}^b}{P_i^b} - 1 - \beta)y + (\frac{P_{i+1}^g}{P_i^g} - 1)g_i + (\frac{P_{i+1}^b}{P_i^b} - 1 )b_i]
    \end{align}
    
    If the investor's utility level is only related to the asset level, then the utility level $U(y_p)$ is also a random variable. From the perspective of maximizing expected utility, the investor's decision-making process is determined by the following formula:
    \begin{align}
        \max_{e_i} E( U(V_0 y_p) ) \sim \max_{e_i} E( U( y_p) )
    \end{align}
    Do the Taylor-expansion on the $E[U(y_p)]$, we can easily find that. if $y_i$ meets normal distribution, the expected utility depends entirely on the mean and variance of the portfolio return, so the investor's decision-making problem turns into:
    \begin{align}
        & \min \sum_{i=1}^n \sum_{j=1}^n e_i e_j \mathbf{Cov(y_i,y_j)} \\
        & \max \sum_{i=1}^n y_i e_i \notag
    \end{align}
    
    Here we assume that cash itself is also an investment option, the only different thing is that its price will not change and cash is independent from the other two assets. We plot the rate of return between September $11{th}$, 2016 and September $10{th}$, 2021:
    \begin{figure}[H]
        \small
        \centering
        \includegraphics[width=0.9\textwidth]{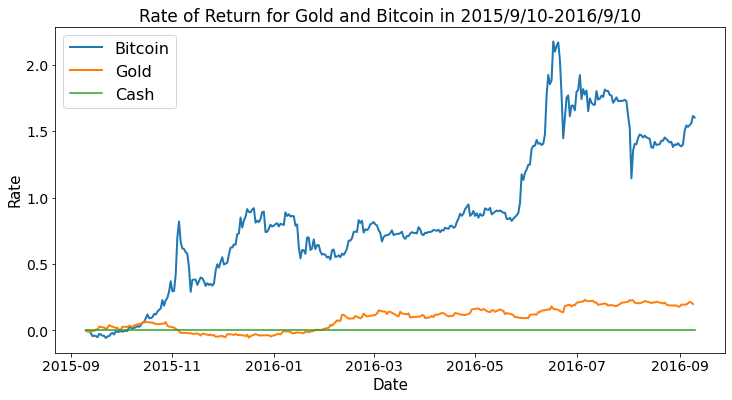}
        \caption{The best time length of $\mathbf{ARIMA}$ model} \label{fig:Rate_GB}
    \end{figure}
    We find that the return rate of bitcoin is much higher than the gold, so we first set $y$ to the maximum value in the process of dynamic programming, and continuously reduce it and increase the value of $x$ to obtain the best investment ratio. And finally the initial \$1,000 dollars on September $10{th}$, 2021 became \$14662.63.Since the above model is based on the assumption that the price fluctuations of bitcoin and gold are normally distributed, which is impossible in our real life. Additionally, we also assumed that gold and bitcoin are relatively independent, but in fact there is a relationship between these two investment products, so we turn to the Sharpe ratio investment model which conforms more to the investment situation.
\subsubsection{Sharpe Ratio Investment Model Based on Particle Swarm Optimization}

     \indent In order to better quantify the relationship between daily investment risk and return, we introduce the Sharpe ratio.Its mathematical expression is:
    \begin{align}
        SR_L=\frac{E(y_L)-y_f}{\sigma_L}
    \end{align}
    in which $E\left(y_L\right),\sigma_L,y_f$ represent the expected return, the standard deviation of the return and the return of the risk-free asset during the observation period respectively.   
    
    Accordinging to the dynamic programming model, on the $i^{th}$ day, we set the target function $\mu(i)$ to be the value of the Sharpe ratio on the third day when the holding position is adjusted the first day and keeps unchanged in the next three days. When the target function reaches its largest number, that is the balance state of risk and gain. Since holding positions unchanged in real life may lead to problems on decisions made everyday, in the case of frequent fluctuations, every time we make a decision, we only consider the situation that the current position condition is maintained to the third day, which may easily lead to selling at a low point and buying at a high point. This obviously will cause unexpected results. So we choose to set the next two days by $\left[x_{i+1},y_{i+1},x_{i+2},y_{i+2}\right]$ value according to the investment model:
    \begin{align}
        \max SR_L
    \end{align}
    as well as the restrictions brought up in the section \ref{The Setup of the model} to solve and come up with the solution.
    
    Due to the large number of decision variables in the investment process, as well as the vague range and the low speed of convergence. We choose to use the particle swarm algorithm to help better speed up the convergence and continue to plan the model under the condition of the initial value which may largely speed up the process of convergence.
    
\begin{figure}[H]
        \small
        \centering
        \includegraphics[width=0.9\textwidth]{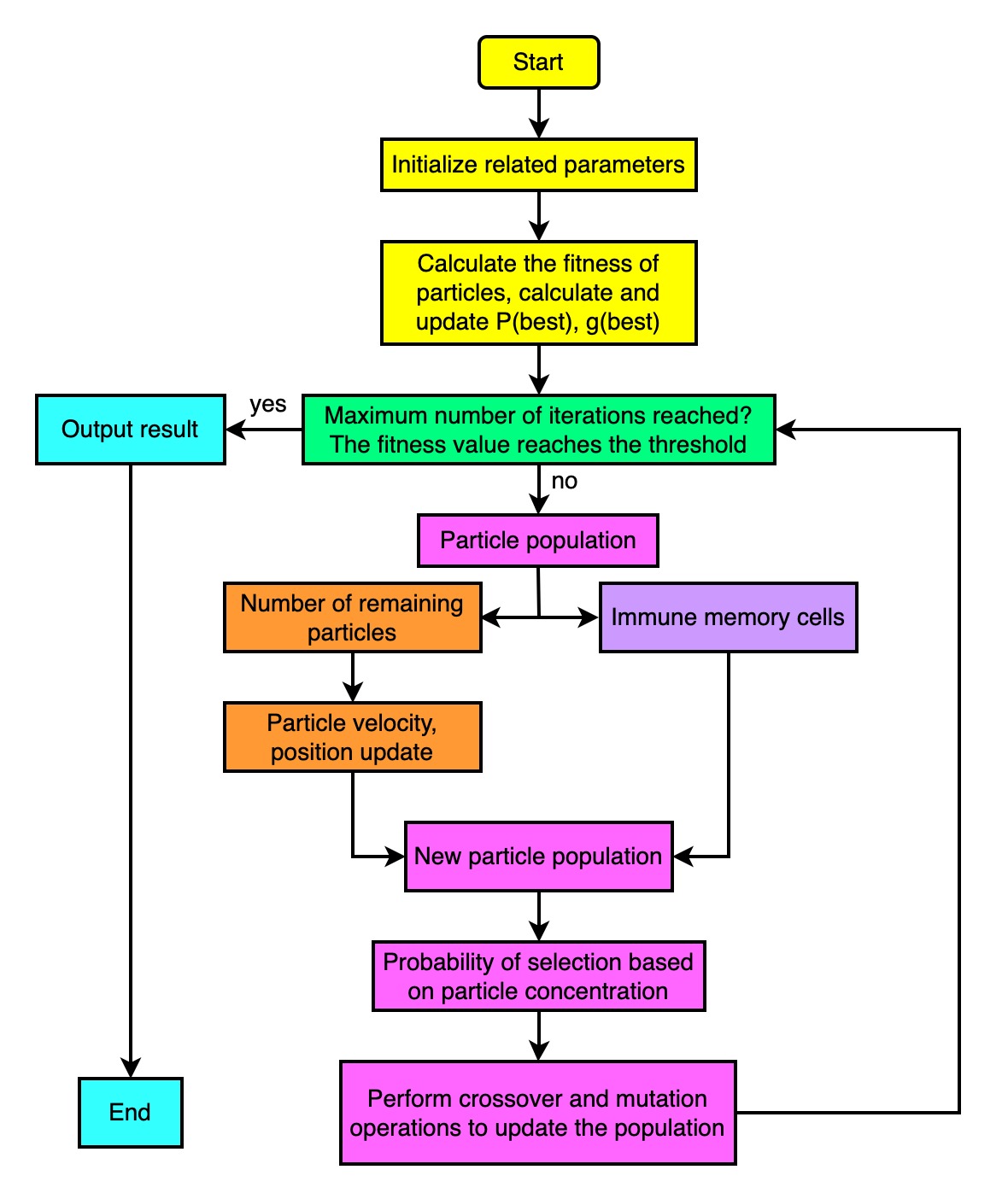}
        \caption{Yield curves of three different investment plans from September $11{th}$, 2016 to September $10{th}$, 2021} \label{fig:pso}
    \end{figure}

    About the Particle Swarm Optimization, which is used above, it is a swarm intelligence algorithm that simulates the mutual cooperation mechanism of the foraging behavior of biological groups in nature to find the optimal solution to the problem. The specific flow chart is shown as figure \ref{fig:pso}
    
    Assuming that in the $N$-dimension space, there are $n$ massless and volumeless particles. Vector $x_i = (x_{i1}\ , x_{i2} ,\dots ,x_{iN} )$ symbolize their position and vector $v_i = (v_{i1} , v_{i2} \dots v_{iN} )$ symbolize their moving speed. In the situation of our problem, on the $i^{th}$ day, we have 6 decision variables in total, which are the $i^{th}$ day, the $i+1^{th}$ day, and the $i+2^{th}$ day's change in gold and bitcoin. That means it is the operation of particle swarm in six-dimensional space, assuming that initially, the number of particles is $100$. Each particle has a moderate value determined by its objective function, which is the objective function and it can find its own optimal position during movement $p_i=(p_{i1},p_{i2},\dots,p_{iN})$, and the optimal position of neighboring particles $g_i=(g_{i1}, g_{i2},\dots,g_{iN})$. Particles get new position parameters by continuously moving and updating iteratively and their velocity and position update formulas are
    \begin{align}
        & v_{ij}(t+1)=\omega v_{ij}(t)+c_1r_1(P_{iN}(t)-x_{iN}(t))+c_2r_2(g_{iN}(t)-x_{iN}(t)) \\
        & x_{ij}(t+1)=\omega v_{ij}(t)+v_{ij}(t+1) \notag
    \end{align}
    In the formula: $\omega$ is the inertia parameter. When $\omega>1$, the particle motion accelerates, and when $\omega<1$, the particle motion decelerates. Now most researchers value $\omega$ from $0.9 $ decreases to $0.4$;$c_1$, $c_2$ are learning parameters, generally take $c_1=c_2 =2$; $r_1, r_2$ are random numbers of $[0,1]$; $x_{ij}( t+1)$ is the position parameter of the particle $i$ in the $j$ dimension iterates $t+1$ times, $v_{ij}(t+1)$ is the particle $i$ iterates in the $j$ dimension Speed parameter for $t+1$ times.
    
    In order to fully consider the impact of investor personality on this model, we specially set risk-return preference $\delta$ and three investment personality types. Risk-return preference represents the minimum proportion of cash to total funds, which means that a constraint must be added in the dynamic programming process.
    \begin{align}
        c_i - x(1 + \alpha) - y(1 + \beta) \geq \delta
    \end{align}

    When simulating the investment character, we do not consider the impact of risk appetite return, even if $\delta=0$, we use the standard deviation $\sigma_i$ of the predicted value in the next three days as the risk assessment value. If the predicted price fluctuation in the next three days is relatively large, also It means that the risk in the future is relatively large. Instead, change the target function.
    \begin{enumerate}
        \item Crazy: Set the target function $\mu(i)$ to be the largest after three days
        \item Stable: Set the target function $\mu(i)$ to the maximum Sharpe rate after three days 
        \item Middle: Set the target function $\mu(i)$ to the return after three days minus 0.618 times the risk assessment value $\sigma_i$
    \end{enumerate}
    
\begin{figure}[H]
        \small
        \centering
        \includegraphics[width=0.9\textwidth]{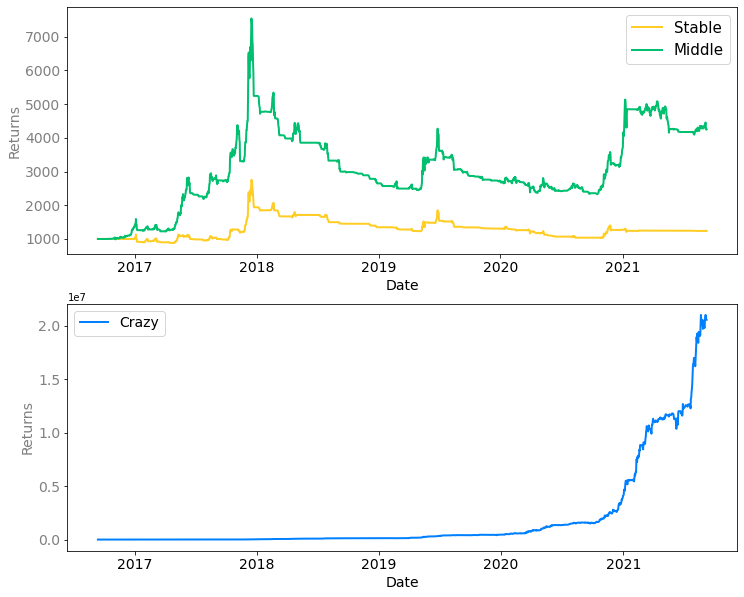}
        \caption{Yield curves of three different investment plans from September $11{th}$, 2016 to September $10{th}$, 2021} \label{fig:portfolio}
    \end{figure}

\begin{figure}[H]
        \small
        \centering
        \includegraphics[width=0.9\textwidth]{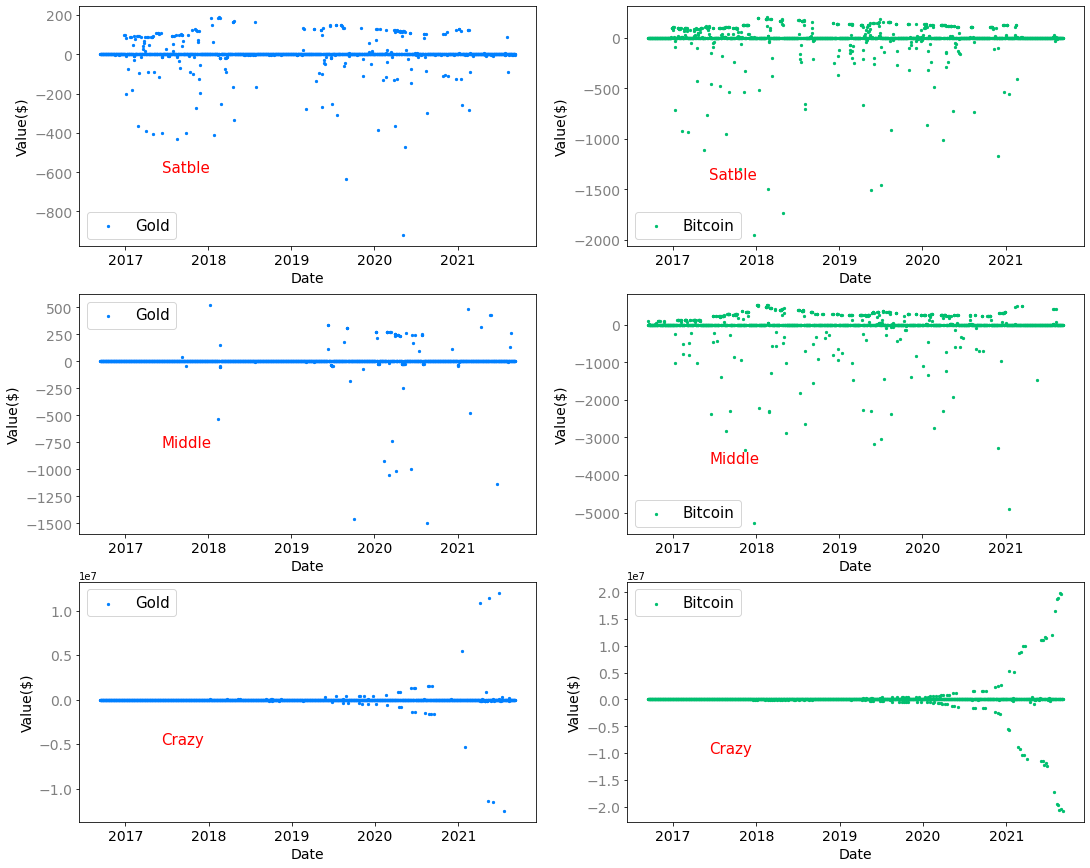}
        \caption{The transaction of gold and bitcoin every day from September $11{th}$, 2016 to September $10{th}$, 2021 in the investment plan} \label{fig:tranctions}
    \end{figure}

     Finally, by September $10{th}$, 2021, the total investment value of the aggressive type is $2.18564 \times 10^7 $, the total investment value of the stable type is 1353.2104, and the total investment value of the intermediate type is 4103.1156. The return curve of each of the three investment schemes As shown in the figure \ref{fig:portfolio}. The transaction of gold and bitcoin every day from September $11{th}$, 2016 to September $10{th}$, 2021 in the investment plan is shown in figure \ref{fig:tranctions}, the point in this figure represents the  daily transaction amount of gold or bitcoin.
\section{Sensitivity Analysis}
\subsection{Prove of the scheme is the best}

    We directly use the mean of all local objective functions $\mu(i)$ as the overall objective function. To prove that the solution we choose is the local optimal solution, we can consider a certain disturbance to the results of our investment plan, and It shows that the result after perturbation is no longer the maximum value of the objective function. Specifically, we added a perturbation in the random range of $1\%-3\%$ to our above-mentioned gold and bitcoin trading scheme, and used the perturbed buying and selling scheme. The results of multiple experiments are as follows:
    \begin{table}[H]
        \centering
        \begin{tabular}{c|cccccc}
            \hline
            Times & $0$ & $1$ & $2$ & $3$ & $4$ & $5$   \\
            \hline
            Middle & $4103.1156$ & $4052.8754$ & $4074.6178$ & $4102.1001$ & $4096.4449$ & $4033.2627$   \\

            Stable & $1353.2104$ & $1121.3841$	& $1289.8158$ & $1275.5071$	& $1310.7616$ & $1303.6777$ 	\\

            Crazy($10^7$) &	$2.1856$ & $2.1806$ & $2.1705$	& $2.1799$ & $2.1840$ & $2.1803$  \\

            \hline
        \end{tabular}
        \caption{Test results after disturbance of investment scheme}
        \label{tab:Test result}
    \end{table}
    It can be seen that the final result after perturbation is not as good as the best result calculated by the model, so the solution obtained by our model is the optimal solution.
\subsection{Sensitivity Analysis}

\begin{figure}[H]
        \small
        \centering
        \includegraphics[width=\textwidth]{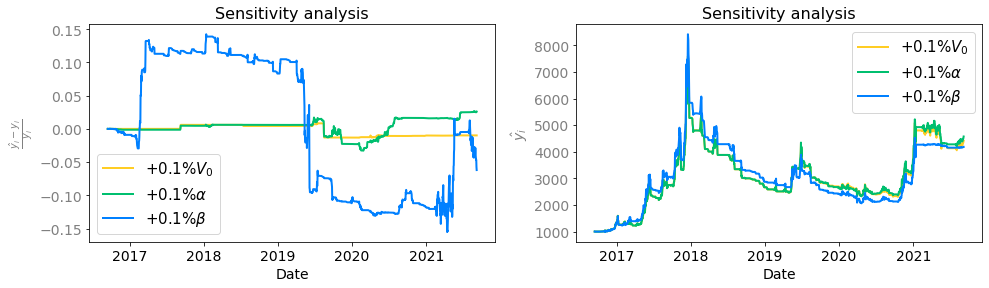}
        \caption{Sensitivity Analysis} \label{fig:sa}
    \end{figure}

    In order to perform a sensitivity analysis of an investment model, we need to start from two angles. In the investment process, the original total assets will have an impact on the future investment plan. Now, in order to examine the strategy of the intermediate personality that we sought before, when the original total assets change ($+0.1\%V_0$), it will generate At the same time, the transaction cost of investment is also very important, so we also adjust the transaction cost $\alpha$ and $\beta$ ($\pm 0.1\%\alpha$ and $\pm 0.1\%\beta$), the above results are shown in the figure \ref{fig:sa} and the table \ref{tab:sa}.
    
    \begin{table}[H]
        \centering
        \begin{tabular}{c|cccccc}
            \hline

            Methods &  $+0\%$ & $+0.1\%V_0$ & $+0.1\%\alpha$ & $+0.1\%\beta$ & $-0.1\%\alpha$ & $-0.1\%\beta$\\
            \hline

            Profit & $4103.1156$ & $4412.9043$ & $4572.5629$ & $4180.4794$ & $4712.8521$ & $5605.5598$\\
            \hline
        \end{tabular}
        \caption{Sensitivity Analysis}
        \label{tab:sa}
    \end{table}
    It can be seen from the results in the figure and table that the model has corresponding changes to the disturbance of the original total assets and transaction costs, that is, it shows the characteristics of "sensitive". At the same time, it can also be seen that transaction costs, especially Bitcoin transaction costs, have a greater impact. When the transaction cost of gold rises or the transaction cost of Bitcoin, the model will tend to increase the investment in Bitcoin. Since the rate of return of Bitcoin is higher than that of gold, the final income will increase; on the contrary, when the transaction cost of gold decreases Or when the transaction cost of Bitcoin rises, the model will reduce the investment options for Bitcoin, making the final return decrease.

\section{Evaluation of Our Model}
\subsection{Advantages}
    \begin{enumerate}
        \item This paper adopts the combination of forecasting model and venture capital model, which ensures the rationality of our investment model.
        \item In this paper, the $\mathbf{ARIMA}$ model combining the optimal time length and the adaptive time length is used for prediction, which effectively reduces the influence of unstable items caused by human factors on the prediction results.
        \item This paper selects two sets of data from 2010 to 2016 Bitcoin price changes and 1966 to 2016 gold price changes to test our model, and the fitting is more accurate.
        \item This article considers the impact of personal personality and other issues on investment, and to a certain extent, various investment scenarios that are in line with our daily life.
    \end{enumerate}
\subsection{Disadvantages}
    \begin{enumerate}
        \item Due to the rapid development of the global economy, the prices of various financial assets, especially Bitcoin, have undergone major changes in recent years. Therefore, the data we select from previous periods often cannot replace the situation in recent years.
        \item The \textbf{ARIMA} model requires that the time series data is stable, or is stable after being differentiated; in essence, it can only capture linear relationships, not nonlinear relationships. Even if we use a combination of optimal time length and adaptive time length, the prediction results are still affected by this unstable nonlinear factor.
        \item The amount of calculation is large. We need to predict the price data after more than 1800 days in September $11{th}$, 2016 to September $10{th}$, 2021, and after the prediction, we need to use the particle swarm algorithm to converge, resulting in our model in the actual operation process. It takes a very long time.
    \end{enumerate}
    
\newpage
\memoto{A trader}
\memofrom{MCM Team 2226491}
\memosubject{Investment scheme}
\memodate{\today}
\begin{memo}[Memorandum]
    We are here writing to offer you a strategy in investment of cash, gold and bitcoin. Based on the price of gold and bitcoin from 2010 to 2017, we develop a model which can determine and describe what transaction you can make on each day and what is the certain maximum return you will get under three different risk levels. \

    \textbf{OUR APPROACH}
    
    To find a optimal strategy of investment, we build the following model:
    \begin{itemize}
    
\item[$\blacktriangleright$]Dynamic Programming Model:By normalized the investment status for each day, we can choose different $\Delta G_i$ and $\Delta B_i$ symbolize different investment strategy.
    
\item[$\blacktriangleright$]prediction Model: we use \textbf{ARIMA} Time Series Model predicts the price change after the investment day, helping us better choose the best investment strategy. On the basis of this model, we also optimize it by a combination of optimal duration and adaptive duration.

\item[$\blacktriangleright$]Sharpe Ratio Investment Model with Particle Swarm Optimization: it would quantify the relationship between investment risk and return on a daily basis better. Taking into account investor personality affects the model differently, We set three investment personality types: Crazy, Stable, Middle.
    \end{itemize}

    \textbf{OUR RECOMMENDED STRATEGY}
    
    From our analysis and results of the model, we conclude three patterns of investing, which is crazy, stable and middle. I will introduce these three modes for you as following:
    \begin{itemize}

\item[$\bigstar$] Crazy: The return graph based on the length of day in investing is shown as below(the third graph) and it can be easily drawn that this mode is more suitable for long-term investing, in which the return may surge on the $5^{th}$ year and the benefits gained may be 20,000 times of the original asset. But such choice may also confront huge risk and rely more on the great fluctuations in the stock market. The first two graph below shows that how you need to invest for gold and bitcoin each day in specific with the blue dots representing gold investment each day and green dots representing bitcoin investment each day.
 \begin{figure}[H]
        \small
        \centering
\includegraphics[width=\textwidth]{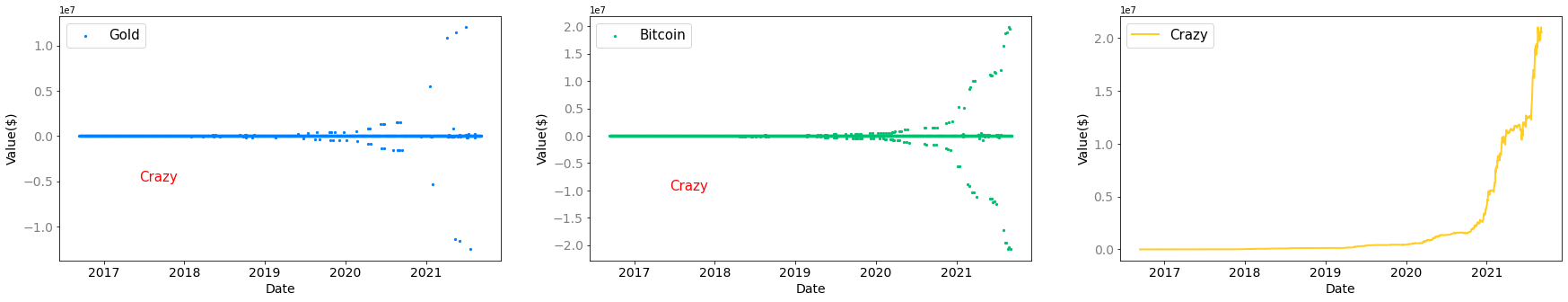}
        \caption{Crazy} \label{fig:crazy}
    \end{figure}

\item[$\bigstar$] Stable: This investing strategy of stable mode may minimize the risk you may take while the benefits in this case is also relatively small. And from the line graph(the third graph) of the returns, you may see that, if you want a round one-year investment and do not want to take little risk, this may be the best choice for you. The investing patterns which describes how you may need to make transactions every day are shown on the first two graph below with the blue dots representing gold investment each day and green dots representing bitcoin investment each day. 
 \begin{figure}[H]
        \small
        \centering
\includegraphics[width=\textwidth]{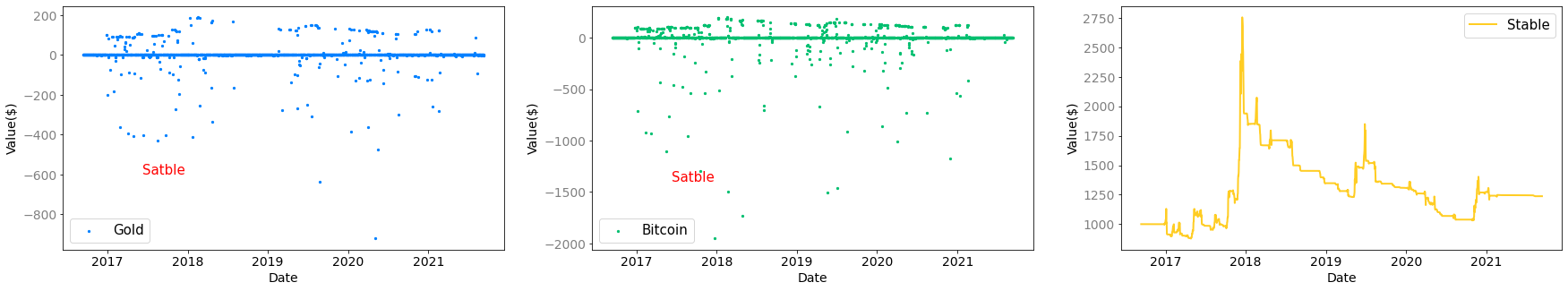}
        \caption{Stable} \label{fig:stable}
    \end{figure}

\item[$\bigstar$] Middle: This investing strategy of middle mode takes both benefits and risks into account, thus we may achieve a balanced state of these two. If you can tolerate certain amount of risk and want to obtain returns more than the stable mode gets, this middle one may be helpful, especially for the short-term investment around one year. More specific investment details are shown on the first two graph below with the blue dots representing gold investment each day and green dots representing bitcoin investment each day.
 \begin{figure}[H]
        \small
        \centering
\includegraphics[width=\textwidth]{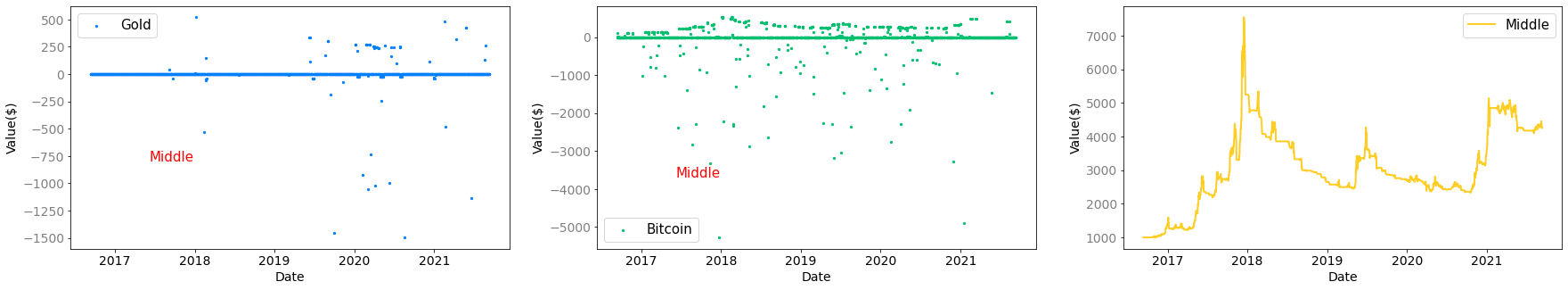}
        \caption{Middle} \label{fig:middle}
    \end{figure}

\end{itemize}
Due to the restrictions of the memo space, we cannot add all the data here. But we can also provide more specific data on the form of table in detail if you want. \\
\rightline{Sincerely,} \\
\rightline{Team 2226491}
\end{memo}

\begin{appendices}

\textcolor[rgb]{0.98,0.00,0.00}{\textbf{ARIMA} code}
\begin{lstlisting}[language=Python]
from statsmodels.tsa.arima_model import ARIMA
import numpy as np, pandas as pd
from statsmodels.graphics.tsaplots import plot_acf, plot_pacf
import matplotlib.pyplot as plt
import pmdarima as pm

df = pd.read_csv('BCHAIN-MKPRU_2010_2018.csv', names=['value'], header=0)
R = []
for i in range(10,11):
    R_temp = 1
    for j in range(i,2219-3):
        model = pm.auto_arima(df[j:j+i].value, start_p=1, start_q=1,
                              information_criterion='aic',
                              test='adf',       
                              max_p=3, max_q=3, 
                              m=1,              
                              d=None,           
                              seasonal=False,   
                              start_P=0, 
                              D=0, 
                              trace=True,
                              error_action='ignore',  
                              suppress_warnings=True, 
                              stepwise=True)

        # Forecast
        n_periods = 3
        fc,confint = model.predict(n_periods=n_periods,return_conf_int=True)
        index_of_fc = np.arange(len(df.value), len(df.value)+n_periods)
        
        # make series for plotting purpose
        fc_series = pd.Series(fc, index=index_of_fc)
        #lower_series = pd.Series(confint[:, 0], index=index_of_fc)
        #upper_series = pd.Series(confint[:, 1], index=index_of_fc)
        Real = [df[j+i+1].value,df[j+i+2].value,df[j+i+3].value]
        Mean = (Real[0] + Real[1] + Real[2])
        Rn = (fc_series[0] - Mean)**2 + (fc_series[1] - Mean)**2 \\
        + (fc_series[2]-Mean)**2
        Rn = Rn / ( (Real[0] - Mean)**2 + (Real[1] - Mean)**2 + \\
        (Real[2]-Mean)**2 )
        R_temp = min(Rn,R_temp)
    R.append[R_temp]

plt.plot(df.value)
plt.plot(fc_series, color='darkgreen')
plt.fill_between(lower_series.index, 
                 lower_series, 
                 upper_series, 
                 color='k', alpha=.15)

plt.title("Final Forecast of WWW Usage")
plt.show()
\end{lstlisting}

\end{appendices}

\begin{thebibliography}{99}
\bibitem{risk:arima model1} Ariyo A A, Adewumi A O, Ayo C K. Stock price prediction using the ARIMA model[C]//2014 UKSim-AMSS 16th International Conference on Computer Modelling and Simulation. IEEE, 2014: 106-112.
\bibitem{risk:omega1} Keating C, Shadwick W F. A universal performance measure[J]. Journal of Performance Measurement, 2002, 6: 59-84.
\bibitem{risk:omega2} Kapsos M, Christofides N, Rustem B. Worst-case robust Omega ratio[J].European Journal of Operational Research, 2014, 234(2): 499–507.
\bibitem{risk:risk1}Olsen R A. Investment risk: The experts' perspective[J]. Financial Analysts Journal, 1997, 53(2): 62-66.
\bibitem{risk:particle swarm algorithm1}Kennedy J, Eberhart R C. A discrete binary version of the particle swarm algorithm[C]//1997 IEEE International conference on systems, man, and cybernetics. Computational cybernetics and simulation. IEEE, 1997, 5: 4104-4108.
\bibitem{risk:particle swarm algorithm2}Shi Y. Particle swarm optimization: developments, applications and resources[C]//Proceedings of the 2001 congress on evolutionary computation (IEEE Cat. No. 01TH8546). IEEE, 2001, 1: 81-86.
\bibitem{risk:risk2}Olsen R A. Investment risk: The experts' perspective[J]. Financial Analysts Journal, 1997, 53(2): 62-66.
\end{thebibliography}
\end{document}